\documentstyle[11pt,fleqn,epsf,epsfig]{article}
\begin{document}
\title{Kinetic Origin of Heredity in a
Replicating System with a Catalytic Network}
\author{Kunihiko Kaneko\\
{\small \sl Department of Pure and Applied Sciences,
College of Arts and Sciences,}\\
{\small \sl University of Tokyo,}\\
{\small \sl Komaba, Meguro-ku, Tokyo 153, Japan}\\
}

\date{}
\maketitle

\begin{abstract}
The origin of heredity is studied as a recursive state in a replicating proto-cell
consisting of many molecule species in mutually catalyzing reaction networks.
Protocells divide when the number of molecules, increasing due to replication,
exceeds a certain threshold.  We study how the chemicals 
in a catalytic network can form 
recursive production states in the presence of errors in the replication process. 
Depending on the balance between  the total number of molecules
in a cell and the number of molecule species, we have found three phases;
a phase without a recursive production state, a phase with itinerancy over a few recursive states,
and a phase with fixed recursive production states.  Heredity is realized in the latter two phases
where molecule species that are population-wise in the minority are preserved and control the
phenotype of the cell.  It is shown that evolvability is realized in the
itinerancy phase, where a change in the number of minority molecules
controls  a change of the chemical state.
\end{abstract}
\vspace{.4in}

Key Words: catalytic network, heredity, minority control, origin of life, evolvability

\pagebreak
\section{Introduction}

In recent studies on isologous diversification\cite{IDT1,IDT2,CF},
we have shown that cell differentiation and developmental processes
are general characteristics of systems of interacting and replicating cells 
that contain mutually catalytic molecules.
In such systems, a cell consists of several chemicals that
catalyze each other. When the chemicals in a cell exceed a
given threshold, the cell divides.  Even though detailed mechanisms are not
programmed in advance, differentiation of cell types, irreversible processes  from 
multipotent
stem cells to commitment, and robust development can be found.

Then one may ask  what the role of genes is.  As a possible answer to this question,
we conjectured that genes come into action to fixate the
the differentiation provided by the interaction of units with
autocatalytic reaction networks.  
The idea is that some of the chemicals of the network provide a hereditary 
starting point that plays a role in controlling the phenotype of the cell.
Indeed, a related idea was
already proposed in the context of evolution, 
by Waddington\cite{Wad} and Newman\cite{Newman}.

Since a gene is a part of the larger structure that is  DNA, 
its action should be represented merely by a
component of the catalytic reactions in the chemical network.
How is it possible then that some chemicals are able to control cell differentiation?

To answer this question, we have recently proposed a minority-control 
theory\cite{MCT}.
In a system of mutually catalyzing molecules,
a state is selected through reproduction where a minor molecule species
starts to control the chemical state of the cell. This state is preserved over many generations of cells.
In this chemical state termed the ``minority controlled state",
a separation of roles between two sets of molecule species appears.
One set has large numbers of molecules, maintaining the diversity of the species,
while the other has small numbers of molecules but 
a larger catalytic activity.
The latter set has 
the following two properties;

[Preservation property]: The molecule species are preserved well over generations.  
The number of molecules per species exhibits smaller fluctuations compared to  
other molecule species, and their chemical structure
(such as polymer sequence) is preserved over a long time span,
even under potential changes caused by thermodynamic fluctuations 
during the synthesis of these molecules.

[Control property]: A structural change in the molecule species or a change 
in the number of the molecules has a stronger effect on the behavior of a
cell, for example on the growth rate of the cell.

These two properties form the basis for heredity, 
causing  statistical correlations between the phenotypes of ancestor and offspring.
Due to the control property, the minority molecules govern the
phenotype, and due to the preservation property, they can transfer the information relevant for
determining the phenotype to future generations.
Indeed, once this minority controlled state is established, a new evolutionary 
incentive emerges
for embedding the representation of heredity into the minority molecule.
Thus the emergence of genetic information from the minority-controlled state is expected.  

In this earlier study, we considered only two molecule species with different
synthesis speeds, one being much higher than the other.  
Or, equivalently, the catalytic activity of the latter is much higher
than the former due to the mutual catalyzation for the synthesis.
The minority controlled state is a consequence of
this difference in the synthesis speeds.
Then, the question remains whether it is a generic property of chemical reaction networks
to select states with large differences in synthesis speeds resulting in
recursive production controlled by a few minority molecule species.
Furthermore, it is necessary to clarify how specific chemical compositions 
are stabilized as recursive states through control by minority molecules.  
Here it is interesting to note that cell differentiation in a model
with interacting cells\cite{IDT1} was initiated by difference in the concentration of
minority molecules by cells.

To answer this question, we consider a protocell system consisting of a large number of molecule
species that catalyze each other.
By carrying out stochastic simulations of such catalytic reaction networks,
we will show in this paper that a biochemical state with minority molecules can 
establish recursive production.  In other words, we will show that the dynamic change of 
the chemical
composition is stabilized by minority molecules that catalyze 
majority molecules, and that the minority molecule species carries this information through
reproduction.
Furthermore, we will also show that this minority controlled state has evolvability,
since by only a slight change in the number of the minority molecules, 
the total chemical composition is drastically altered.

\section{Model}

To study the general features of a system with mutually catalyzing molecules\cite{Eigen},
we consider the following simple model.  First, we envision
a (proto)cell containing $k$ molecule species.  (Not all of these species
are necessarily present. Some of the molecule species may have zero population).
With a supply of
chemicals available to the cell, the molecules replicate through catalytic
reactions, so that their numbers within a cell increase.  When the total number
of molecules exceeds a given threshold (here we used 2$N$), the cell divides into two,
with each daughter cell inheriting half of the molecules of the mother cell, 
chosen randomly. 
Each
chemical species catalyzes the synthesis of some other randomly chosen chemical
as

\begin{equation}
X^i + X^j \rightarrow 2X^i + X^j.
\end{equation}

\noindent
with $i,j=1,\cdots,k$.  The connection rate of the catalytic path is given by $p$
(which will be taken to be 0.2 or 0.1 for most simulations here), and the
connection is chosen randomly. (Here we investigated the case without direct mutual
 connections, i.e.,  $i\rightarrow j$ was excluded as a possibility when there 
was a path $j \rightarrow i$, although this condition is not essential for the results
 to be discussed). Each molecule species $i$ has a given 
catalytic ability $c(i)$ and its own synthesis speed $g(i)$.  Accordingly, the
above reaction occurs with the rate $g(i)c(j)$. 

During the replication process
structural changes may occur that alter the activity of the molecules. Therefore the
replicated molecule species can differ from that of the mother.  
The rate of such structural changes is given by $\mu$, which may not
be very small due to thermodynamic fluctuations.
This change can consist of the alternation of a sequence in a polymer or other conformational
change, and may be regarded as a replication `error'.
Here, for the simplest case, we take
this `error' to affect all molecule
species equally, (i.e., with the rate $\mu /(k-1)$), and could thus 
regard it as a background fluctuation.
Generally in reality of course, even after a structural change, the replicated molecule
would keep some similarity with 
the original molecule, and the replicated species with the `error' would be
within a limited class of molecule species.
Later we will discuss this case also, but the basic conclusion will not 
be altered. Hence we use the simplest case for most simulations.


We simulated this model according to the following procedure.
In the beginning, we choose connection matrix randomly.  Unless otherwise mentioned,
the parameters $c(i)$ and $g(i)$ are given from a random number over $[0,0.5]$,
and they are fixed through each simulation.
At each step, a pair of molecules, whose species  is assumed to be  
$i$ and $j$, is chosen randomly.  If there is a connection between the species $i$ and $j$, then 
according to the reaction (1), one molecule of the species $i$ is added
with the probability $g(i)c(j)$ if $j$ catalyzes the synthesis of $i$,
or $j$ is added with the probability $g(j)c(i)$ if $i$ catalyzes the 
synthesis of $j$.
The replications are subjected to errors with the rate $\mu$ given above.
The error rate $\mu$ is fixed at 0.01 through the simulations here, but
this specific choice is not so important as long as it is not too large.
Since our model deals with rather small numbers of molecules, a stochastic approach was
chosen for the simulations rather than the usual  ordinary differential equations for 
chemical concentrations which cannot describe some novel effects \cite{Togashi}.

When the number of molecules within a cell is larger than $2N$, 
it is divided into two, and a new cell is created.
The total number of cells, $M_{tot}$, is kept  constant  
so that one protocell, randomly chosen,
is removed whenever a (different) protocell divides into two.
Often we use $M_{tot}=1$ here, where only one of the daughter cells remains.
In this case, our model is quite similar to a stochastic simulation of the population
dynamics of $k$ molecule species, but at each division event
large fluctuations arise.  Hence a recursive state that continues production 
has to be selected, resulting in a choice of special initial conditions for the
chemical composition.  This selection of initial conditions through repetition of
divisions is important, especially when $N$ is not large (or $k$ is large).
This selection of initial conditions distinguishes our model
from standard population dynamics of molecules.
With this model we address the question how recursive production is achieved, and 
how evolutionary change of the recursive state is possible.

\section{Recursive States and Itinerancy}

In our model there are four basic parameters; the total number of molecules 
$N$, the total number of molecule species $k$, the mutation rate $\mu$,
and the path rate $p$, besides the total cell number $M_{tot}$, 
the catalytic activity $c(i)$ and growth rate  $g(i)$.
By investigating many sets of parameters and also by 
choosing various random networks, we have found 
that there are roughly three types of behaviors:

(1) Fast switching states without recursiveness

(2) Itinerancy over several recursive states

(3) Achievement of fixed recursive states

In the first phase, there is no clear recursive production and the dominant
molecule species changes frequently.  At one time
step, some chemicals may be dominant but only a few
generations later, this information is lost, and the number of the
molecules in this species goes to zero.  
Due to the autocatalytic nature, the 
population of one species can be amplified, only to be replaced by another population
catalyzed by it.  No stable set of
catalytic networks is formed that excludes other `parasitic' molecules.
(See for example the dynamics of chemicals around the last stage of Fig.1 a).
In phase (1), this type of dynamics continues forever,
without showing  any  quiescent state as around the middle of Fig.1a)).

\begin{figure}
\noindent
\vspace{-.1in}
\hspace{-.3in}
\epsfig{file=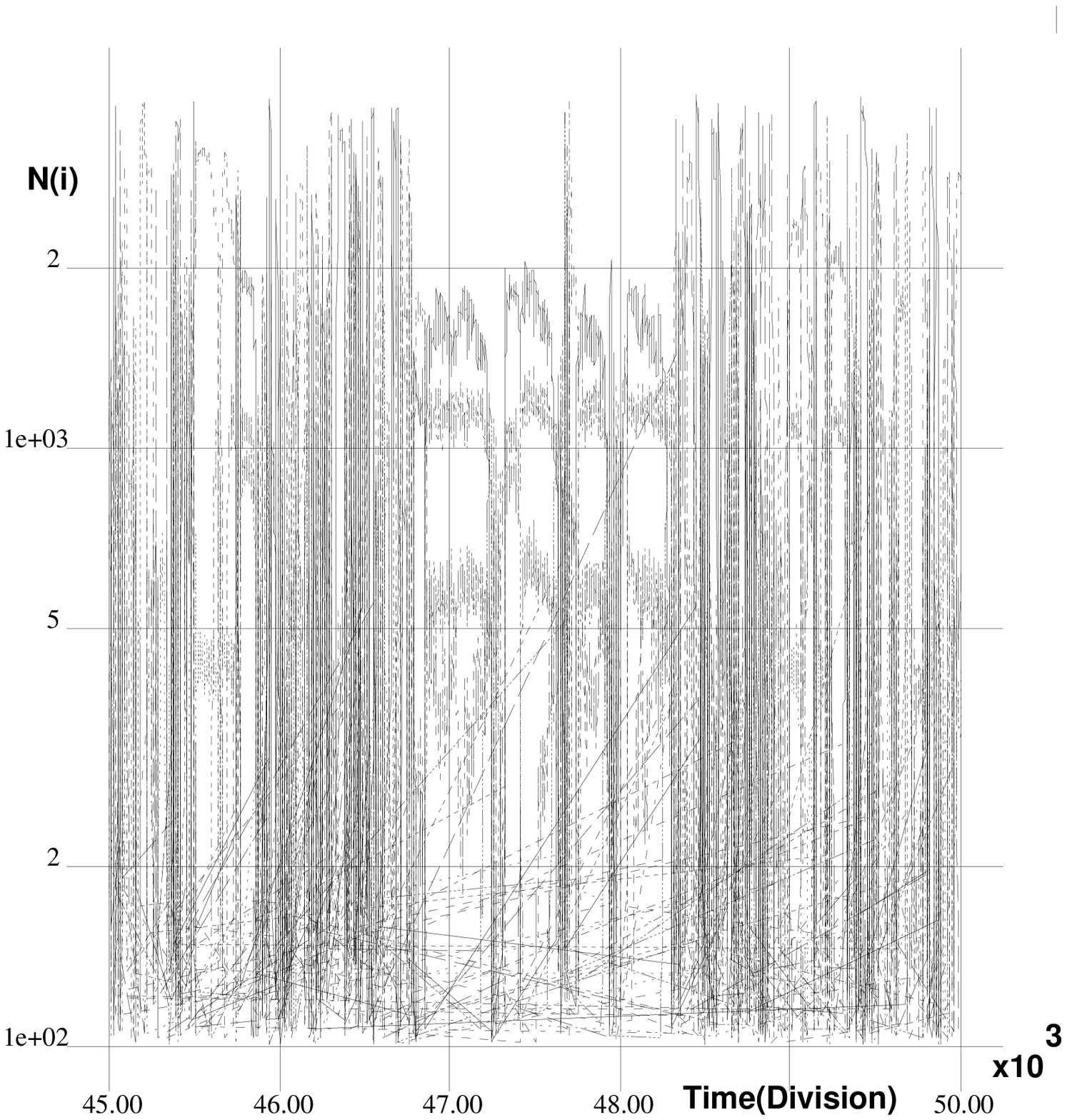,width=.6\textwidth}
\epsfig{file=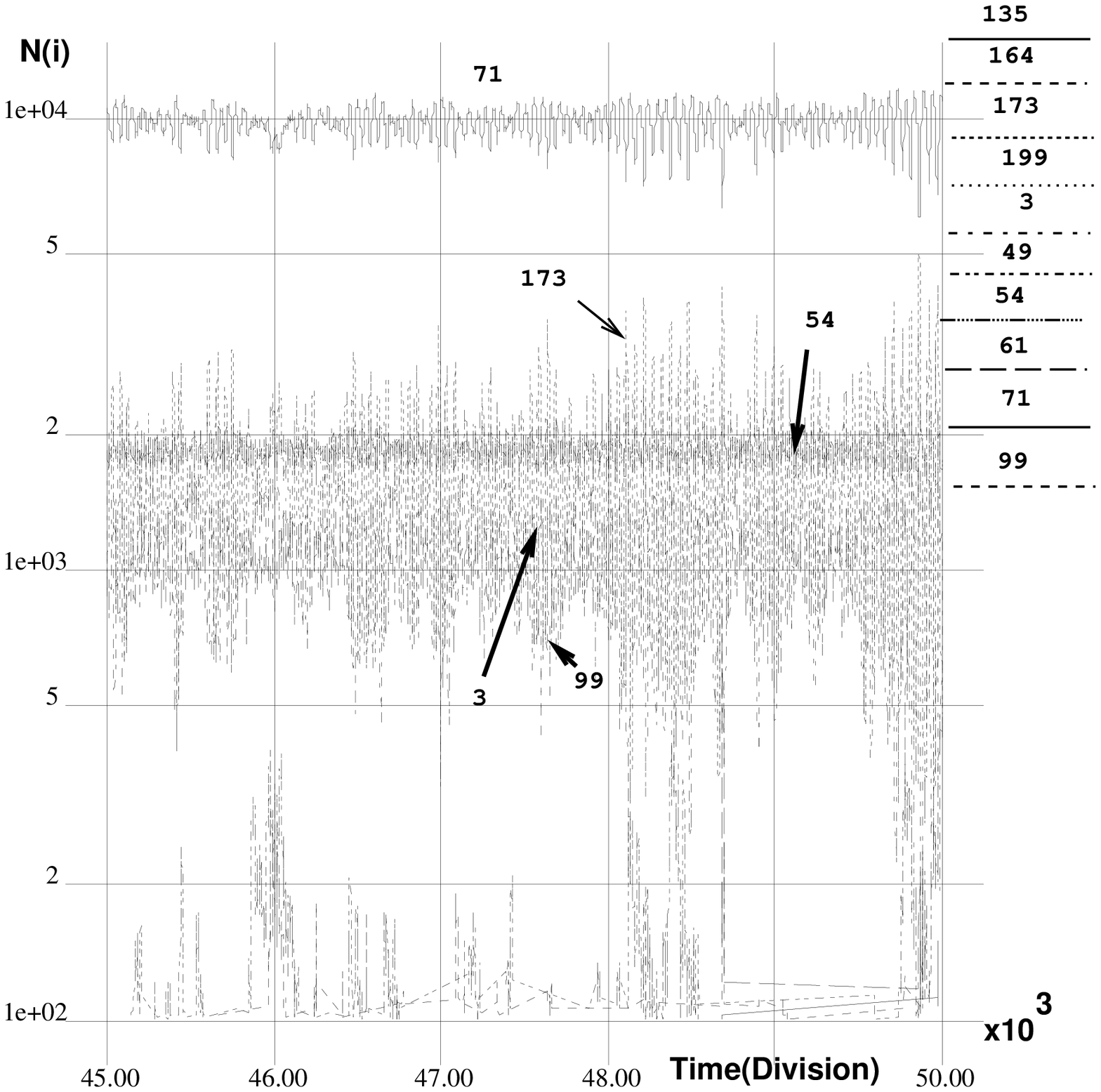,width=.6\textwidth}
\caption{The number of molecules $N(i)$ for the
species $i$ is plotted at each successive division event. 
In both (a) and (b), the same random network with $k=200$ and $p=.2$ was used while $N=2000$ in (a), 
and $N=6000$ in (b).  Only the species with $N(i)>10$ are plotted at each time.
The straight lines are artifacts of the drawing occurring when $N(i)$ dips below 10 and then at a
later time increases to above 10 again. 
The parameters $c(i)$ and $g(i)$ are randomly distributed over [0,0.5] and are fixed
through the simulation. In (a), there appears a recursive state around 
$45.8\times 10^3$-th generation, and another recursive state appears around
$47\times 10^3$-th generation and repeats collapse and reappearance up to $48.3\times 10^3$-th generation. In (b), a single recursive state lasts
through the whole simulation, where 5 species (71,173, 3, 54, 99
in the order of population size) continue to exist.}
\end{figure}

In the second phase, alternately, 
recursive states which last over many generations (typically a thousand generations)
and fast switching states appear  (See Fig. 1 a) 
\footnote{Such switching between
several recursive states is also studied as chaotic itinerancy \cite{CI1,CI2}}.

In the third phase, on the other hand, a recursive state is established,
and the chemical composition is stabilized such that it is not altered 
by the division process. Once reached, this state is permanent and the system remains in it 
for as long the simulation lasts. (see Fig.1b).

To be precise, the recursive state observed in phases (2) and (3) is not necessarily a fixed point 
as with regards to the population dynamics of the chemical concentrations.
In some cases, as shown in Fig.1b), the chemical concentrations 
oscillate in time, but the nature of the
oscillation is not altered by the process of cell division.
In dynamical systems terms, the recursive state here
is an attractor, and this attractor can be a limit cycle (or could be 
chaos or some other
low-dimensional attractors), even though often it actually is a fixed point 
attractor.  (In all of these cases there are fluctuations due to the
small number of molecules, of course).
Generally, all the observed recursive states consist of
5-10 species, except for those species which exist only as a result of replication errors
where the number of molecules is one or two. 

Which of the states (1), (2), (3) appears depends, of course, on the parameters and the 
specific choice of the network.
First, we discuss how the phase changes when varying $N$ and $k$ if 
$c(i)$ and $g(i)$ are distributed homogeneously over
$[0,0.5]$, $M_{tot}=1$, and $p=.2$, since  
we have not yet elucidated any clear dependence on $p$ and $M_{tot}$ so far.

Although the behavior of the system depends on the choice of the network,
there is a general trend with regards to the phase change,
from (1), to (2), and then to (3) with the increase of  $N$, 
or with the decrease of $k$, as is schematically drawn in Fig.2.
 
\begin{figure}
\noindent
\hspace{-.3in}
\epsfig{file=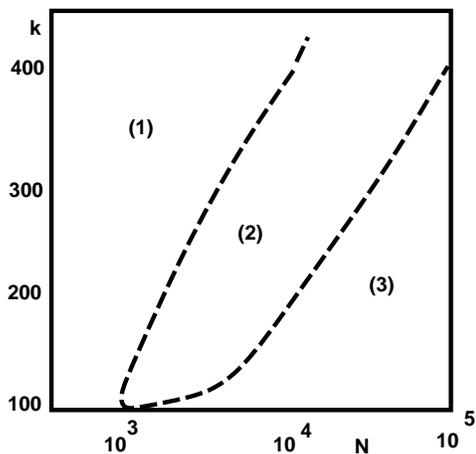,width=.5\textwidth}
\caption{Schematic phase diagram,
plotted as a function of the total number of molecules $N$,
and the total possible number of molecule species $k$.}
\end{figure}

We now study what role the catalytic activity plays in maintaining the recursive states.
In Fig.3, we have plotted the population  of each molecule species as a function of
the catalytic activity.  For phase (1), no clear structure is discernible, although
a slight tendency for the average population to decrease with an increase in
the catalytic activity exits as is  
shown in  Fig.3a).  The maximum of each population, however, 
depends only minimally on the catalytic activity, implying that almost all
species can be the dominant species at some time.

For phase (2), a clear structure can be observed and the maximum populations decrease 
for increasing catalytic activity.  
Indeed, this negative correlation between catalytic activity and population 
is not surprising.  If the catalytic activity is higher it can
help the synthesis of other molecules.  Since the total number of molecules is limited, 
the fraction of that molecule population itself should be less.

\begin{figure}
\noindent
\hspace{-.3in}
\epsfig{file=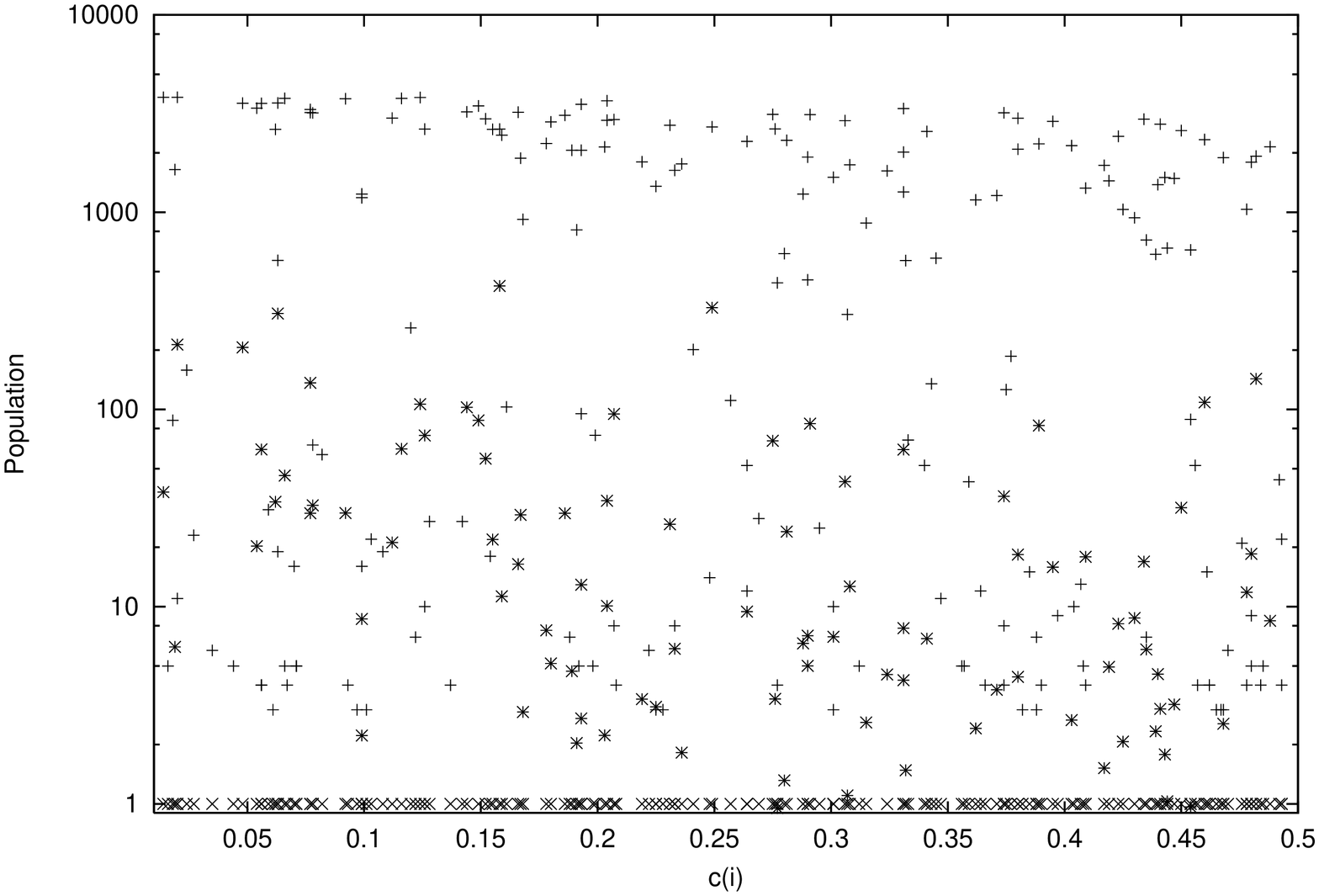,width=.6\textwidth,angle=-90}
\epsfig{file=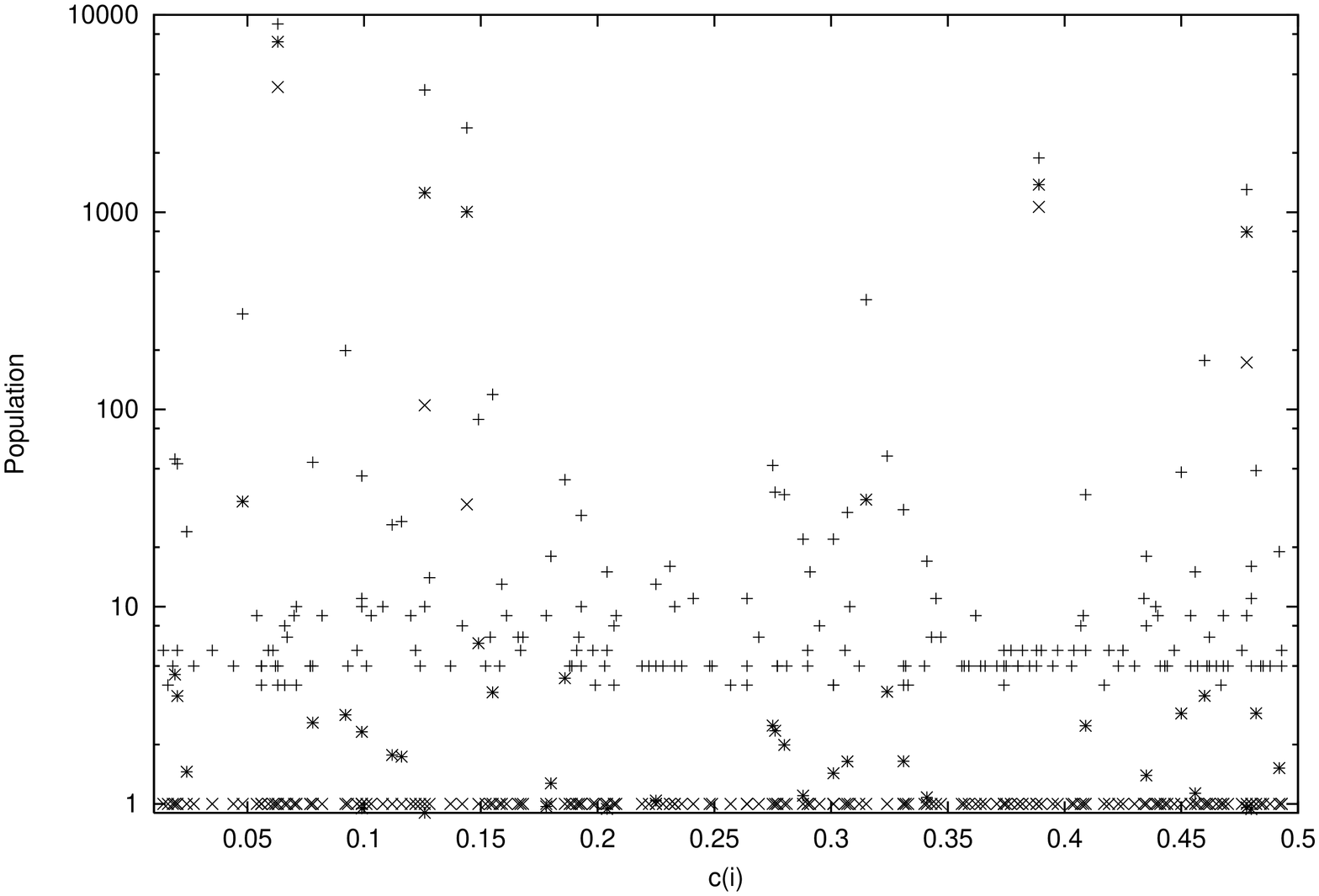,width=.6\textwidth,angle=-90}
\caption{Average($*$), maximal(+) and minimal($\times$) populations of the 
species $i$, plotted as a function of their catalytic activity $c(i)$.
The average, maximum and minimum are taken over the 50000-100000th division.
In (b), the five dominant species are discernible, which are species
71, 173, 3, 54, 99 in the order of population size (i.e., in the reverse order
of $c(i)$).}
\end{figure}

This negative correlation is amplified and fixed in phase (3), as shown
in Fig. 3b).  
Here, only a few molecule species with small populations and high catalytic activity survive.
Although these molecules are in the 
minority population-wise, their existence is essential for maintaining this recursive
structure.  The major species have  lower catalytic activities, and accordingly they are not
catalyzed efficiently by the other major molecule species.
Indeed, the species with the majority of the population are catalyzed by the
minority species.  The latter is catalyzed by species with larger populations
but with smaller catalytic activities.  The surviving molecule species form a 
mutually catalyzing reaction network validating the assumptions made
in the minority control theory.

For example, in the recursive state in Fig.1b), there are 5 species
whose population remains in existence for all the time.  The network structure 
of these molecule species is shown in Fig.4, with other molecules that exist 
not over all the time but over relatively long time span.
The catalytic activities of the key species are $c(99)=.48$$>c(54)=.39$$>c(71)=.063$, 
while the respective populations are
$N(71) \gg N(54) \gg N(99)$.  The recursive state here is 
achieved by catalysis of the species 99 whose population is
a minority in the network.
When the total population is much smaller, the species $99$ may
 decrease due to  fluctuations and competition with other
molecules catalyzed by the species 71.  For example, in the collapse
of the recursive state in Fig.2a), the population of the
species $3$ increases due to  catalyzation by $71$, and a state dominated by $71$, $54$, 
and $3$ is formed.  Since the catalytic activity of these molecules is not 
so high, the state is dynamically unstable and allows for an increase
in the populations of other molecules.  
For example, the increase of the species $61$ leads to 
the increase of species $18$ and $137$ $\cdots$, successively, and the
recursive state is replaced by the fast switching state.

By examining several reaction networks, we came to understand itinerancy and 
the stability of recursive states as follows.  
As for the recursive state,
one might wonder why the species with higher catalytic activities 
are not taken over by parasitic molecule species that have
lower catalytic activities and are catalyzed
by molecules with higher catalytic activities.  
To sustain a recursive state,
the emergence of such parasitic molecule species should be suppressed. 
In our model, molecules with higher catalytic activities are catalyzed by a 
molecule species with lower activities but larger populations.
Hence, the parasitic molecule species cannot easily invade to
disrupt the mutually catalytic core network.
Since the minority molecule (say $99$ in the above example) and
majority molecule ($71$ in the above example) form a mutual catalytic network
(with the aid of another molecule ($54$)), a large fluctuation
is required to destroy this mutually supporting network.

\begin{figure}
\noindent
\hspace{-.3in}
\epsfig{file=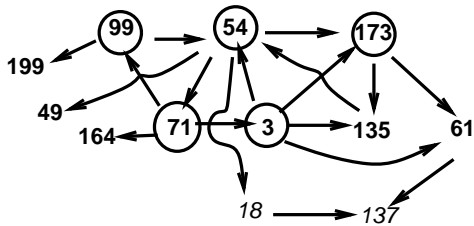,width=.5\textwidth}
\caption{The core network for the recursive state (with numbers in bold), 
and a part of the parasitic molecules (with numbers in italic).}
\end{figure}

For recursive stability, it is important that there are two levels of replication,
i.e., molecular replication and cellular replication\cite{Szath}.
As parasitic molecules with lower catalytic activity appear,
recursiveness is lost, and the state starts to change, while
the total growth speed of the cell generally decreases.  With  cellular  
replication, the selection of initial conditions maintaining the recursive state
is thus favored.
In the minority controlled state, minority and majority molecules catalyze and reinforce
each other, and the recursive state is stabilized.

This allows us to explain how the itinerancy over (partially) recursive states
appears in phase (2).  When the number of molecules is not so large, the numbers of
some molecule species that are not in the original core network start to increase.
Since the total number of molecules in a cell is  limited,  the minority molecules may then
decrease in number, and, if the total number of molecules is small enough, even
become extinct.  If that happens, the molecule species with a large population loses the 
main source that catalyzes it.
Hence several molecules mostly of lower catalytic activity start to compete for
growth.  Consequently, the diversity of molecule species increases, a
stable recursive production cycle of cells is not established, and   
the dominant species change frequently.  
After the system has been in a fast switching state for a while,  
another ( or possibly the same) core network structure
is formed where the existence of a minority molecule species 
with a higher catalytic activity stabilizes the recursive state.

With the above considerations is it quite reasonable to expect
the transition from phase (2) to (3) for increasing $N$. After all, 
for larger $N$, the number of molecules in the minority species also increases
thus protecting them from the extinction.
However, the transition from (2) to (3) also strongly depends on the network structure.  
For some reaction networks, a simple core structure is easily formed
and the recursive state is maintained even for small $N$.  For some other networks,
the itinerancy of phase (2) is observed for large $N$.
Although it is not so easy to distinguish these two 
cases by just examining the network structure, some differences seem to exist in
the dynamics of the recursive state.  If the state falls on a 
fixed point,
 recursiveness is stabilized for smaller $N$.  On the
other hand, when the population shows oscillatory behavior, 
dynamic instability is generated, and fluctuations
in the species with small molecule numbers are continuously amplified. 
Hence the state is subjected to
larger overall perturbations and stabilized only at larger $N$. 

Here the distribution of catalytic activity is important.  We studied
the cases that either $c(i)$ or $g(i)$ or both are homogeneous.
When $c(i)$ is constant, the itinerancy state of phase (2) is rarely observed.
The itinerancy state is common, however, when the distribution of $c(i)$ is
biased to having more species with lower catalytic activity.  
For example, we carried out simulations with $\log{c(i)}$ homogeneously distributed.
In this case, the itinerancy over recursive states is a common occurrence as, 
for example, can be seen in Fig.5.
Considering that enzymes with higher catalytic activity are rare,  this
choice of distribution of $c(i)$ with a bias towards a zero value is rather natural. 
Hence, the itinerancy over
recursive states should be rather common in general.


\begin{figure}
\noindent 
\hspace{-.3in}
\epsfig{file=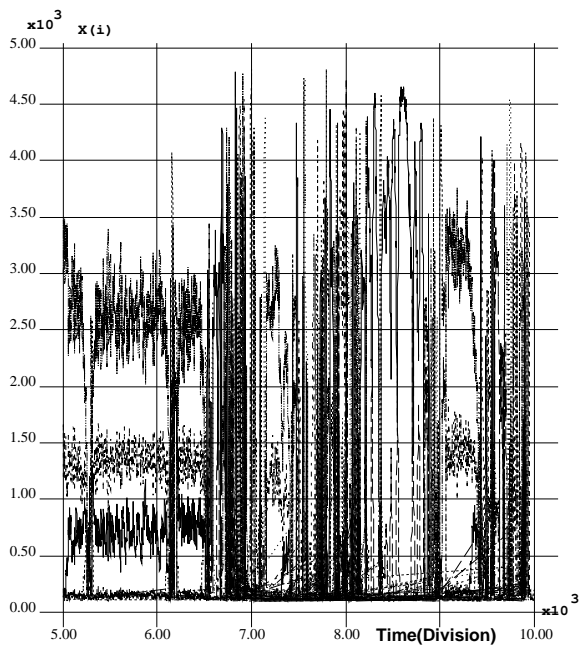,width=.6\textwidth}
\caption{An example of an overlaid time series for the case with a
biased distribution of $c(i)$.  $N=2000$ and $k=200$.}
\end{figure}
 

\section{Discussion: Remark on Evolvability}

An important consequence of the above itinerancy state is evolvability (see also
\cite{MCT}).
With a change 
in the number of minority molecules, or by a structural change in one of them,
the chemical composition of a cell can change drastically and  establish a novel
recursive state\footnote{See \cite{IDT1} and \cite{Ike} for the relevance of minority species 
in a cell differentiation model and in some ecological networks, respectively}.  To investigate 
this evolvability, we have modified the
mutation condition such that one molecule species
can only mutate into a limited range of other molecule species, i.e., 
a species $i$ can only mutate to a species in the range $[i-m,i+m]$ with $m<k/2$.
Starting from an initial condition consisting only of species 
within a range $[k_1,k_2]$, one can then examine whether the 
number of molecule species  expands to the whole range $[0,k]$.
We found that in phase (2), the evolutionary process
`scans' over all the  molecule species (i.e. $i \in [0,k]$),
by  itinerating over several recursive states.
This is a typical feature of the second phase.  For phase (3),
the chemical state is stuck in the recursive state formed at an
early stage, while in phase (1), no recursive state exists.

A recursive state in a mutually catalytic system was also discussed by
Lancet\cite{Lancet} as a `compositional genome'. Recursive states are furthermore similar to
the determined states in the cell 
differentiation model \cite{IDT1,CF} where the chemical composition of a cell is
transferred to its offspring.
There, stable states are understood as attractors or as  partial attractors
stabilized by the interaction.  The recursive state resulting from
such a partial attractor is stabilized by a single molecule species in the current model.

This recursive state forms a few discrete states among a huge variety of possible
chemical compositions, as an attractor. This state gives a basis for heredity.
In genomic information, on the other hand,
each chemical state of a cell is represented by a single molecule.
To bridge the gap between the attractor viewpoint
and information viewpoint, each attractor state has to be represented by
a single molecule and has to be stabilized.  This bridge is provided by
the minority controlled theory we proposed.
Besides recursiveness, evolvability is an important consequence
of the minority controlled state.

The minority-controlled state will generally be important for understanding the 
`bottle-neck' phenomena
ubiquitous in biological systems.
Often a process in a cell includes some checkpoints, and 
each checkpoint has to be passed for a cellular process 
to continue to the next stage.
For example, consider  the timing of a cell division.  To start the division,
a cell has to complete several processes.  Then the slowest dynamics
involved in the cell division process forms a  bottleneck.  With such a bottleneck,
a recursive cellular state is guaranteed.  (For example, most of the gadgets 
in the cell, even though some of them require a longer time for synthesis, 
are doubled before the division).  With the
minority-controlled state, a dynamics to overcome the bottleneck is provided.
Since the minority molecule
is necessary for the catalysis of many other molecules, the cell has to wait for it to be replicated
before it proceeds to the next stage.  If many minority molecules are combined together as a
single long polymer sequence, then this polymer will act as
the control of all the other cell processes.  Probably this is the role of DNA,
and the origin of this fascinating information carrier.

\vspace{.1in}

{\bf Acknowledgment}

The authors would like to thank Walter Fontana, Tetsuya Yomo, Shin'ichi Sasa and Takashi Ikegami
for useful discussions and Frederick Willeboordse for critical reading of the manuscript.
This research was supported by Grants-in-Aid for Scientific Research from
the Ministry of Education, Science and Culture of Japan (11CE2006).

\vspace{.1in}




\end {document}